# Large-Scale Integrated Vector-Matrix Multiplication Processor Based on Monolayer MoS$_2$

Guilherme Migliato Marega[1,2], Hyun Goo Ji[1,2], Zhenyu Wang[1,2], Mukesh Tripathi[1,2], Aleksandra Radenovic[3], Andras Kis[1,2*]

[1]Institute of Electrical and Microengineering, École Polytechnique Fédérale de Lausanne (EPFL), CH-1015 Lausanne, Switzerland
[2]Institute of Materials Science and Engineering, École Polytechnique Fédérale de Lausanne (EPFL), CH-1015 Lausanne, Switzerland
[3]Institute of Bioengineering, École Polytechnique Fédérale de Lausanne (EPFL), CH-1015 Lausanne, Switzerland
*Correspondence should be addressed to: Andras Kis, andras.kis@epfl.ch

**Led by the rise of the internet of things, the world is experiencing exponential growth of generated data. Data-driven algorithms such as signal processing and artificial neural networks are required to process and extract meaningful information from it. They are, however, seriously limited by the traditional von-Neuman architecture with physical separation between processing and memory, motivating the development of in-memory computing. This emerging architecture is gaining attention by promising more energy-efficient computing on edge devices. In the past few years, two-dimensional materials have entered the field as a material platform suitable for realizing efficient memory elements for in-memory architectures. Here, we report a large-scale integrated 32×32 vector-matrix multiplier with 1024 floating-gate field-effect transistors (FGFET) that use monolayer MoS$_2$ as the channel material. In our wafer-scale fabrication process, we achieve a high yield and low device-to-device variability, which are prerequisites for practical applications. A statistical analysis shows the potential for multilevel and analog storage with a single programming pulse, allowing our accelerator to be programmed using an efficient open-loop programming scheme. Next, we demonstrate reliable, discrete signal processing in a highly parallel manner. Our findings set the grounds for creating the next generation of in-memory processors and neural network accelerators**

**that can take advantage of the full benefits of semiconducting van der Waals materials for non-von Neuman computing.**

Over the past decade, billions of sensors from connected devices have been used to translate physical signals and information to the digital world. Due to their limited computing power, sensors integrated into embedded remote devices often transmit raw and unprocessed data to their hosts. However, the high energy cost of wireless data transmission[1] affects device autonomy and data transmission bandwidth. Improving their energy efficiency would open a new range of applications while reducing the environmental footprint. This motivates the desire to shift data processing from remote hosts to local sensor nodes so that data transmission would be limited to structured and valuable data. In this context, the von-Neuman architecture, with its separation of memory and logic units, is widely seen as the most critical limiting factor for the efficiency of computing systems in general and edge-based devices in particular. The separation between the processing and memory imposed by the von Neumann architecture requires that the data be sent back and forth between the two during data and signal processing or inference in neural networks. This intense data communication between the memory and the processing unit already accounts for a third of the energy spent in scientific computing[2].

The desire to overcome the Von-Neumann communication bottleneck[3,4] motivates the rise of in-memory computing architectures in which memory, logic and processing operations are collocated. Such processing-in-memory devices are especially suitable for performing vector-matrix multiplication, which is the key operation for data processing and the most intensive calculation for implementing machine-learning algorithms. By taking advantage of the memory's physical layer to perform the multiply and accumulate operation (MAC), this architecture overcomes the Von-Neumann communication bottleneck. So far, this processing strategy has shown promise for applications such as solving linear[5,6], and differential equations[7], signal and image processing[8], and in artificial neural network accelerators[9–12]. However, the search for the ultimate material and device for realizing this type of processor is still ongoing. Several devices have been studied for in-memory computing, from resistive random access memories (RRAM) to ferroelectric memories (FeFET)[3,13–16]. More recently, two-dimensional materials have shown promise in the field of beyond-CMOS devices[17–22] and in-memory and in-sensor computing[23–26]. Floating-gate field-effect transistors (FGFET) based on monolayer $MoS_2$ have been shown to be scalable[25,27,28]. They can be used for logic-in-memory[29] or in-memory computing, building perceptron layers. Here, they are projected to offer more than an order of magnitude improvements in power efficiency compared to CMOS-based circuits[28]. Even though these past realizations have sparked interest and highlighted the promise of two-dimensional materials for in-memory computing, further progress and real-world applications require wafer-scale fabrication and large or very-large system integration. Currently, demonstrations of wafer-scale and integration of 2D semiconducting-based circuits have been limited to photodetectors[30–33] or traditional analog and digital integrated circuits[34–38]. However, full-wafer and large-system integration involving 2D-based non-volatile memories that can perform general-purpose computation are missing. The realization of such a system would allow in-memory processors to reap all the benefits of 2D materials for the next generation of in-memory processors and open the way to realizing non-Von Neumann computing systems based on 2D materials.

To bring this next generation of in-memory processors closer to reality, we demonstrate a chip containing a 32×32 floating-gate field-effect transistor matrix with 1024 memory devices per chip and an 83.1% yield (please refer to Supplementary information for more details). The working devices show a similar $I_{DS}$ versus $V_G$ hysteresis and characterization behavior. During the fabrication, we use wafer-scale metalorganic chemical vapor deposited (MOCVD) monolayer $MoS_2$ as the channel material, and the entire fabrication process is carried out in a 4-inch line cleanroom. We further demonstrate multi-bit data storage in each device with a single programming pulse, allowing us to overcome the need to use write-verifying schemes, making the programming considerably faster. Finally, we show that our devices can be employed in the context of in-memory computing by performing discrete signal processing with different kernels in a highly parallelized manner.

**Memory Matrix**

Here, we approach in-memory computing by exploiting charge-based memories using monolayer MoS$_2$ as a channel material. Specifically, we fabricated floating-gate field-effect transistors (FGFET) to take advantage of the electrostatic sensitivity of 2D semiconductors[17]. To enable the realization of larger arrays, we organized our FGFETs in a matrix in which we can address individual memory elements by carefully choosing the corresponding row and column. Figures 1a and b show the three-dimensional rendering of the memory matrix and the detailed structure of each FGFET, respectively. The use of a matrix configuration allows a denser topology and corresponds directly to performing vector-matrix multiplications. Our memories are controlled by local 2nm/40nm Cr/Pt gates fabricated in a gate-first approach. This allows us to improve the growth of the dielectric by atomic layer deposition[34] and minimize the number of processing steps that the 2D channel is exposed to, resulting in an improved yield. The floating gate is a 5 nm Pt layer sandwiched between 30 nm HfO$_2$ (block oxide) and 7 nm HfO$_2$ (tunnel oxide). Next, we etch vias on the HfO$_2$ to electrically connect the bottom (M1) and top metal (M2) layers. This is required for routing the source and drain signals without an overlap. Wafer-scale MOCVD-grown MoS$_2$ is transferred on top of the gate stack and etched to form the transistors' channels. Details about material quality and characterization can be found in the Supplementary Information. Finally, 2nm/60nm Ti/Au is patterned and evaporated on top, forming the transistors' drain-source contacts as well as the second metal layer. Further details about the fabrication can be found in the methods section and in the Supplementary Information. Figure 1c shows the optical image of the fabricated chip containing 32 rows and 32 columns a total of 1024 memories. In the image, source channels are accessed on the bottom, the drain channels from the right, and gate channels from the left.

Our memories are based on standard flash memories. The memory mechanism relies on shifting the neutral threshold voltage ($V_{TH0}$) by changing the number of charges in the trapping layer ($\Delta Q$), i.e., the platinum floating gate in our case. When a high positive/negative bias is applied to the gate, the band alignment starts favoring the tunneling in/out of electrons from the semiconductor to the floating gate, changing the carrier concentration in the trapping layer. We define our memory window ($\Delta V_{TH}$) by taking the difference between the threshold voltage from the forward and reverse paths, which are taken at a constant current level. Our previous work verified the programming mechanism by fitting our experimental curves in a device simulation model[25,27]. Since the memory effect relies entirely on a charge-based process, flash memories tend to have better reliability and reproducibility than emerging memories that are material dependent such as resistive random-access memories (RRAM) and phase change memories (PCMs)[3]. We designed and manufactured a custom device interface board (DIB) to facilitate the characterization of the memory array, with a detailed description in Supplementary Information. Figure 1d shows the $I_{DS}$ versus $V_G$ sweeps performed for each device. The fabrication presents a yield of 83.1% and good reliability and reproducibility. The relatively high OFF-state current is due to a lack of resolution of the analog to digital converters used in the setup. High-resolution single-device measurements confirm typical OFF-state currents on the order of pA. Figure 1e shows the ON and OFF current distribution over the memory matrix. Both ON and OFF currents are taken at $V_{DS} = 100mV$ creating 2 distinct planes. The ON and OFF current shows a good distribution over the entire matrix. Further detailed single-device characterization can be found in the supplementary information, confirming the performance of the devices as memories with good retention and endurance stabilities. We show that the devices have a statistically similar memory window $\Delta V_{TH} = 4.30 \pm 0.25$ V. This value is smaller compared to the the one extracted from single-device measurement due to the higher slew-rates (5 V/s) required for time-effective charaterisation of 1024 devices in the matrix.

**Open-Loop Programming**

The similarity of the devices motivates us to pursue a statistical study of the memories' programming behavior. In the context of in-memory computing, an open-loop programming analysis is fundamental. Standard write-verify approaches may be too time-consuming while programming a large flash memory array. A statistical understanding of memory states in open-loop is essential to improving performance and speed.

We perform the experiment such that each device is independently excited by selecting the corresponding row (i) and column (j). Analog switches in the device interface board keep a low impedance path in the selected row (i) / column (j) and high impedance in the remaining rows and columns (Supplementary Information). This ensures that a potential difference is only applied to the desired device, avoiding unwanted programming. For the same reason, we divide the device programming and reading into two independent stages. During the programming phase, the corresponding gate line (row) and the corresponding source line (column) are selected and programming pulses with parameters $T_{PULSE}$ and $V_{PULSE}$ are applied in the gate. Due to the tunneling nature of the device, only two terminals are required to generate the band bending needed for the charge injection into the floating gate. After the pulse, the gate voltage is changed to $V_{READ}$, which is low enough to prevent reprogramming the memory state. In the reading phase, the drain line is also connected, and the conductance value is probed by applying a voltage $V_{DS}$ in the drain. This two-stage procedure is required because we are using a 3-terminal device; therefore, both gate and drain share the same row, and consequently, the entire row is biased when the gate and drain line is engaged. If high voltages in the gate were applied when the drain line is connected, the whole row would be reprogrammed, causing the loss of information in the memories. Figure 2a shows the description of this two-stage programming procedure.

For the subsequent measurements, we used $V_{READ} = -3$ V, $V_{DS} = 1$ V, and $T_{PULSE} = 100$ ms. Before each measurement, we reset the memories by applying a positive 10 V pulse which puts the devices into a low conductance state. Due to parasitic resistances in the matrix, a linear compensation in the digital gains is applied (see Supplementary Information for further details). The compensation method improves the programming reliability of the devices by an order of magnitude. We estimate a programming error of 500 errors per million for programming 1-bit while having 1 error per million for programming the erase state. Figure 2b, c shows the distribution of memory states after different pulse intensities, $V_{PULSE} = +10V, -4V, -6V, -8V$, and -10V in both linear and logarithmic representations. We observe that on a linear scale, the increase in the pulse amplitude is accompanied by a higher memory state value and a larger spread. On the other hand, by analyzing the logarithm of the state value, we can see that the memory has well-defined defined storage states. This leads us to conclude that this memory has the potential for reliable and scalable multivalued storage without write-verify algorithms at a decent programming error.

Figure 2d shows the spatial distribution of the states on the entire chip. We observe that the memory states create a constant plane value for the different programming voltages, $V_{PULSE}$. Finally, Figure 2e shows the empirical cumulative distribution function (ECDF) of the logarithmic representation. These results support the possibility of multivalued programming, as discussed previously and indicate that the memory elements can be used for storing analog weights for in-memory computing.

**States and Vector-Matrix Multiplications**

With the open-loop analysis completed, in Figure 3a, we plot the memory states (<w>) as a function of the programming voltage ($V_{PROG}$). We define four equally distributed states (2-bit resolution) to be programmed as discrete weights in the matrix for the vector-matrix multiplication (please refer to Supplementary information for more details). To analyze the effectiveness of the processor for performing vector-matrix operations, we compare, in Figure 3b, the normalized theoretical ($y_{THEORY}$) value with the normalized experimental ($y_{EXP}$) value obtained on several dot-product operations. The linear regression of the experimental points shows a line with parameters **a** = 0.988±0.008 and **b** = -0.129±0.003 for $y_{EXP}$ = **a**.$y_{THEORY}$ + **b**, while the shaded area corresponds to a 95% confidence interval. The ideal processor should converge to **a** = 1 and **b** = 0 with a confidence interval that converges to the linear fitting. In our case, the processor has a linear behavior converging to the ideal case, with a large spread and slightly non-linearity of the experimental values. We explain this behavior by the non-ideality of the memories and the quantization error due to the limited resolution of the states. The shift in parameter **b** can be explained by a non-perfect OFF state of the memories seen at $y_{THEORY}$ = 0, but it does not affect the observed linear trend. We conclude that we can perform multiplication-accumulation operations with reasonable accuracy. This operation is needed for performing diverse types of algorithms, such as signal processing and inference in artificial neural networks.

**Signal Processing**

Next, we configure this accelerator to perform signal processing to demonstrate a real-world scenario and application. For signal processing, the input signal (x) is convoluted with a kernel (h) resulting in the processed signal (y). Depending on the nature of the kernel elements, different types of processing can be achieved. Here, we limit ourselves to 3 different kernels that perform respectively low-pass filtering, high-pass filtering, and feedthrough. All the kernels run in parallel within a single processing cycle, demonstrating the efficiency of this processor targeting data-centric problems by parallelized processing. Figure 4a shows the convolution operation and the different kernels used for processing the input signal. The strategy to encode negative kernel values into the memories conductance values is to split the kernel (h) into a kernel with only the positive values ($h^+$) and one with the absolute values of the negative numbers ($h^-$) and encode only the positive numbers with a direct relation with the conductance values (G). After the processing is realized, the outputs of the positive ($y^+$) and negative ($y^-$) kernels are subtracted ($y^+ - y^-$), resulting in the final signal (y).

Figure 4b shows the comparison between the original weights and the weights transferred into the memory matrix using the previously described open-loop programming scheme. To simplify the transfer, we normalize the weight values at each kernel. As a result, we observe a good agreement between the original and experimental values. Next, to verify the effectiveness of the processing, we first construct our input signal (x) as a sum the sinusoidal waves with different frequencies. In this way, we can easily probe the behavior of the filters at different frequencies without creating an overly complex signal. Since the signal has positive and negative values, the signal amplitude must fall on the linear region of the device operation. Thus, we restrict the signal range from -100 mV to 100 mV at $V_{READ} = 0$. Figure 4c shows the fast Fourier transform of simulated processed signals on the left and the experimental signals on the right. The grey line in both simulated and measured signals is the fast Fourier transform of each kernel, giving a guideline for the predicted behavior of each operation. We highlight that the experimental processing of all three filters matches quite well the theoretical values as well as the prototype filter.

Here, we have demonstrated large-scale integration of 2D materials as the semiconducting channel in an in-memory processor. We demonstrated the reliability and reproducibility of our devices both in terms of characterization and the statistical similarity of the programming states in the open-loop programming. The processor carries out vector-matrix multiplications and demonstrates its functionality by performing discrete signal processing. This functionality and integration represent a milestone for in-memory computing, allowing in-memory processors to reap all the benefits of 2D materials and bringing new functionality to edge devices for the internet of things.

## ACKNOWLEDGMENTS

We thank Z. Benes (CMI) for help with electron-beam lithography. We acknowledge support from the European Union's Horizon 2020 research and innovation program under grant agreements 829035 (QUEFORMAL), 785219, and 881603 (Graphene Flagship Core 2 and Core 3), 964735 (EXTREME-IR) from the H2020 European Research Council (ERC, grants no 682332 and 899775) as well as from the CCMX Materials Challenge grant 'Large area growth of 2D materials for device integration'. Device preparation was carried out in part in the EPFL Centre of MicroNanotechnology (CMI).

## METHODS

**Wafer Scale Memory Fabrication**

The fabrication starts with a silicon substrate with a 270 nm thick $SiO_2$ insulating layer. The first metal layer and FGFET gates were fabricated by photolithography using an MLA150 advanced maskless aligner with a bilayer LOR 5A/AZ 1512 resist. The 2 nm/40 nm Cr/Pt gate metals were evaporated using an e-beam evaporator under a high vacuum. After resist removal by dimethyl sulfoxide (DMSO), DI water and $O_2$ plasma are used to further clean and activate the surface for $HfO_2$ deposition. The blocking oxide is deposited by thermal atomic layer deposition using TEMAH and water as precursors. The 5 nm Pt floating gates were patterned by photolithography and deposited using the same process as described previously. With the same atomic layer deposition system, we deposit the 7 nm tunnel oxide layer. After the transfer of $MoS_2$ onto the substrate, patterning it with photolithography and etching by $O_2$ plasma.

Drain-source electrodes are patterned by photolithography and 2 nm/60 nm Ti/Au is evaporated in the same machine. To increase the adhesion of contacts and the $MoS_2$ onto the substrate, a 200 °C annealing step is performed in high vacuum. The devices have a W/L ratio of 49.5 μm/ 3.1 μm.

**Device Passivation**

The fabricated device is the first wire-bonded onto a 145-pin PGA chip carrier. The device is heated inside an Ar glovebox at 135°C for 12 hours which removes the adsorbed water from the device surface. After the in-situ annealing in the glovebox, a lid is glued onto the chip carrier using a high-vacuum epoxy and cured in an argon atmosphere. This protects the device from oxygen and water.

**Transfer procedure**

The MOCVD-grown material is first spin coated with PMMA A2 at 1500 rpm for 60 s and baked at 180 °C for 5 min. Next, we attach a 135 °C thermal-release tape onto the $MoS_2$ sample and detach it from the sapphire in deionized water. After this, we dry the film and transfer it onto the patterned substrate. Next, we bake the stack at 55 °C for 1 hour. We remove the thermal release tape by heating it on the hot plate at 130 °C. Next, we immerse the sample in an acetone bath for cleaning the tape polymer residues. Finally, we transfer the wafer to an isopropanol bath and dry it in the air.

**MOCVD Growth**

Monolayer $MoS_2$ was grown using the MOCVD method. $Mo(CO)_6$, $Na_2MoO_4$ and diethyl sulfide (DES) were used as precursors. NaCl was spin-coated as a catalyst. Pre-annealed 3-inch c-plane sapphire wafer with a small off-cut angle (< 0.2°) was used as a growth substrate (UniversityWafer Inc.). The CVD reaction was performed using a home-built furnace system with 4-inch quartz tube reactor and mass flow controllers connected with Ar, $H_2$, $O_2$, and metalorganic precursors (Mo(CO)$_6$ and DES). For the MoS$_2$ crystal growth, a reactor was heated to 870 °C at ambient pressure for 20 minutes.

**Electrical Measurements**

The electrical measurements were performed using a custom device interface board connected to a CompactRIO (cRIO-9056) running a Real-Time LabVIEW server. We have the modules NI-9264 (16 channels analog output), NI-9205 (32 channels analog inputs), and NI-9403 (Digital IO) installed.

## AUTHOR CONTRIBUTIONS

A.K. initiated and supervised the project. G.M.M. fabricated the devices, designed/prepared the measurement setup, and performed the device characterization and remaining measurements. H.G. and Z.W. grew the two-dimensional material and assisted in material characterization under supervision by A.R. M.T. and performed HRTEM for characterization of devices and materials. A.K. and G.M.M. analyzed the data. The manuscript was written by A.K., G.M.M. with the input of all authors.

## COMPETING FINANCIAL INTERESTS

The authors declare no competing financial interests.

## DATA AVAILABILITY

The data that support the findings of this study are available in Zenodo at http://dx.doi.org/XXXX.

# FIGURES

## a    FGFET Matrix

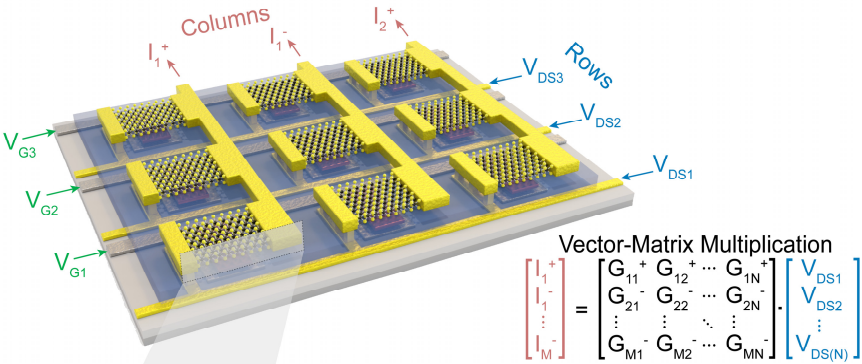

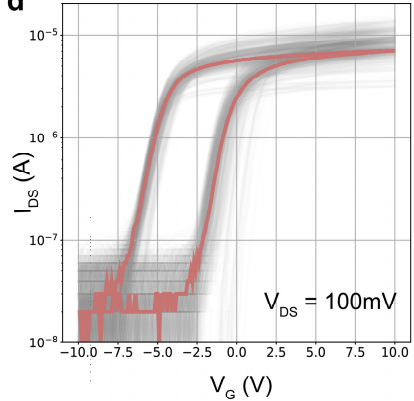

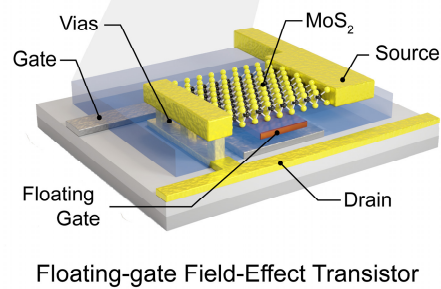

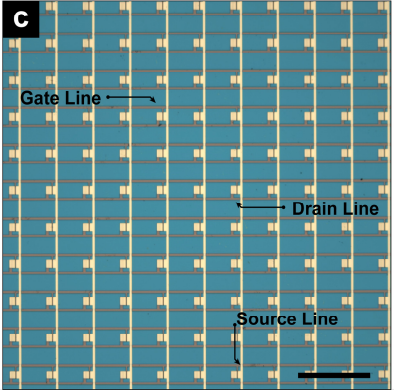

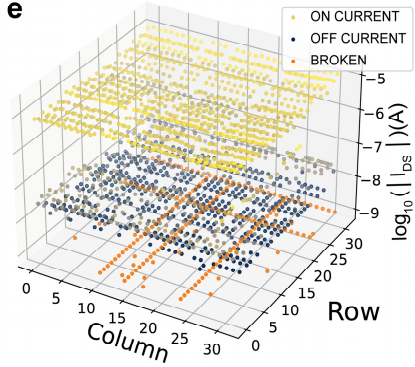

**Figure 1. Device and matrix description and characterization**. **a,** 3D rendering of the floating-gate field-effect transistors connected into a matrix array. Both gate and drain contacts are organized in rows and the source signal is applied to the columns. Gate signals are applied on the left side while drain signals on the right. The drain-source current is read into the column. The inset shows the correspondence between signals and vector-matrix multiplication. **b,** 3D rendering of the floating-gate field-effect transistor cross section. It shows the different device parts. **c,** Optical image of the memory matrix configuration. Scale bar: 500 µm **d,** $I_{DS}$ versus $V_G$ hysteresis curves of the 851 working devices. **e,** 3D plot shows the mapping of the ON current and the OFF current. Devices in orange are disconnected.

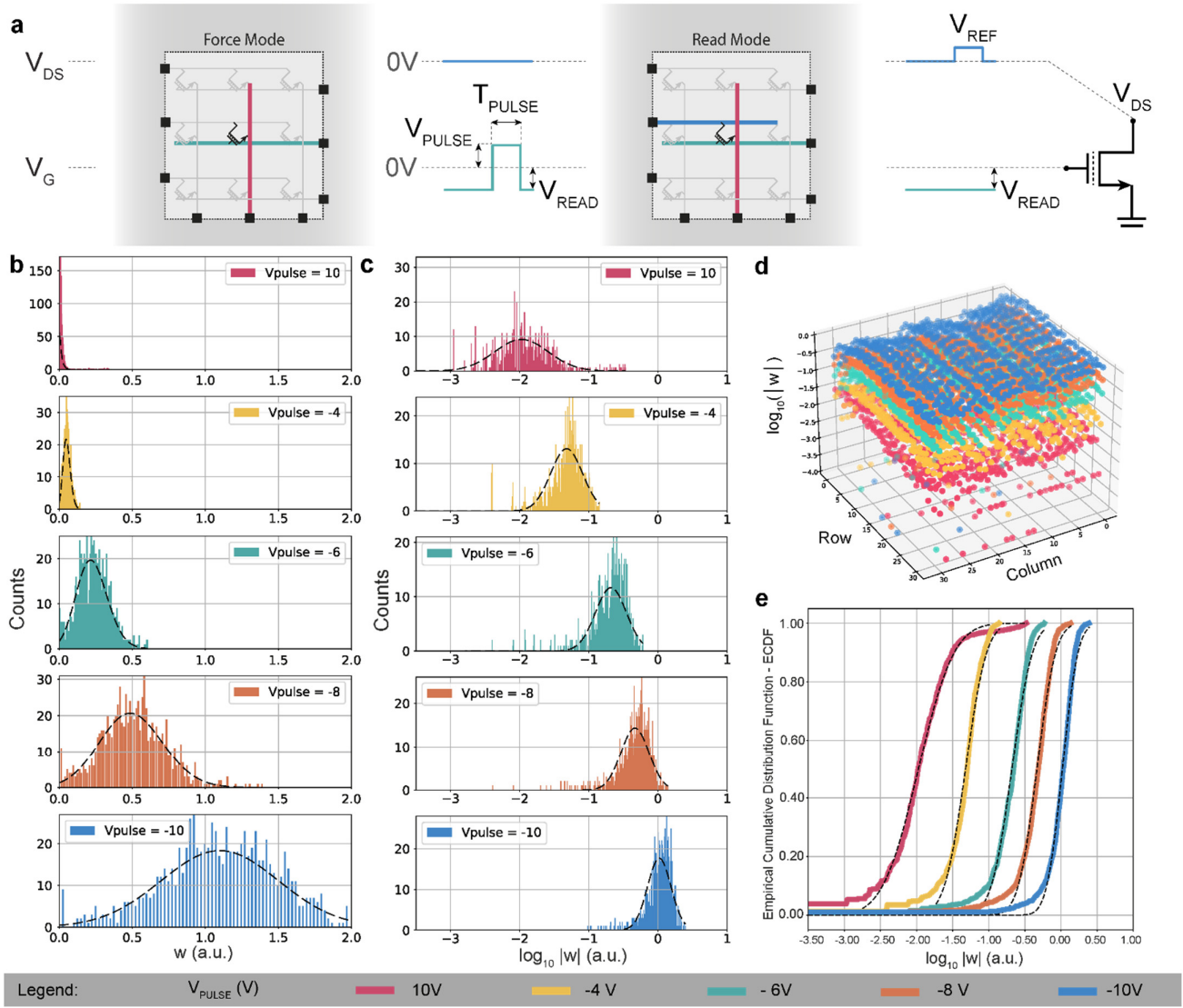

**Figure 2. Open-loop programming a,** Schematic of the 2-state operation of the open-loop programming scheme. In the programming phase, the interface board is used to set the gate line and the source line to low impedance and the drain line to high impedance states, while in the reading phase, all three lines are set to the low impedance state. **b**, Distribution of output states ($w_{OUT}$) in linear scale. The data is fitted with a gamma distribution. **c,** Distribution of output states ($w_{OUT}$) in log10 scale. The distributions are fitted with a gaussian distribution. **d,** 3D map of $\log_{10}$ of the $w_{OUT}$ as a function of device position and different programming voltages. **e,** Empirical cumulative distribution function (ECDF) as a function of the programmed states in $\log_{10}$ scale.

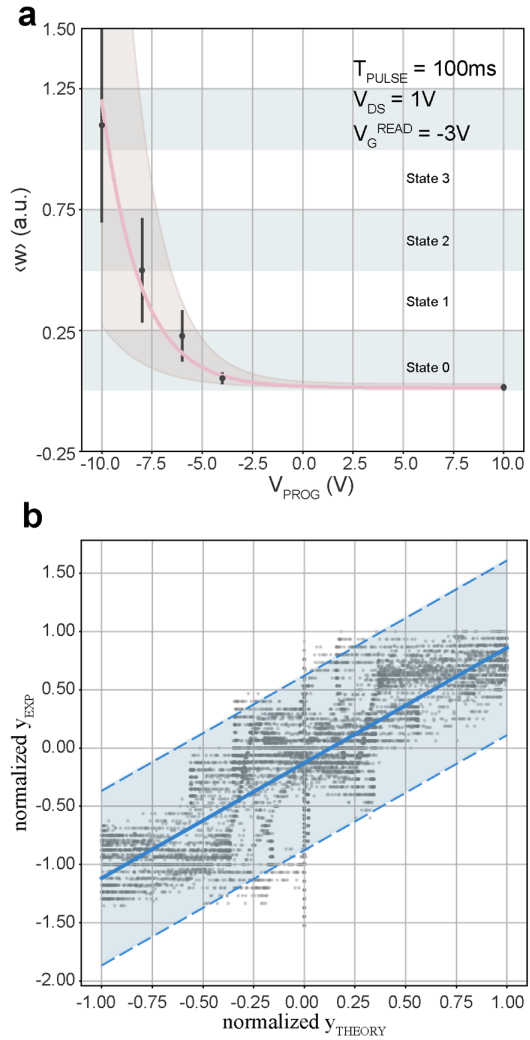

**Figure 3. Multiplication-accumulation operations a,** Output memory states with programming error (<w>) as a function of programming voltage ($V_{PROG}$). To define the state positions, we perform a fit and select the corresponding state branches for a 2-bit open loop operation **b,** Normalized $Y_{EXP}$ versus $Y_{THEORY}$ plot comparing experimental the theoretical results of the multiplication-accumulation operation. The curve is fitted with a linear function with parameters a = 0.988±0.008 and b = -0.129±0.003. The shaded area corresponds to the interval with 95% of confidence of the linear fitting.

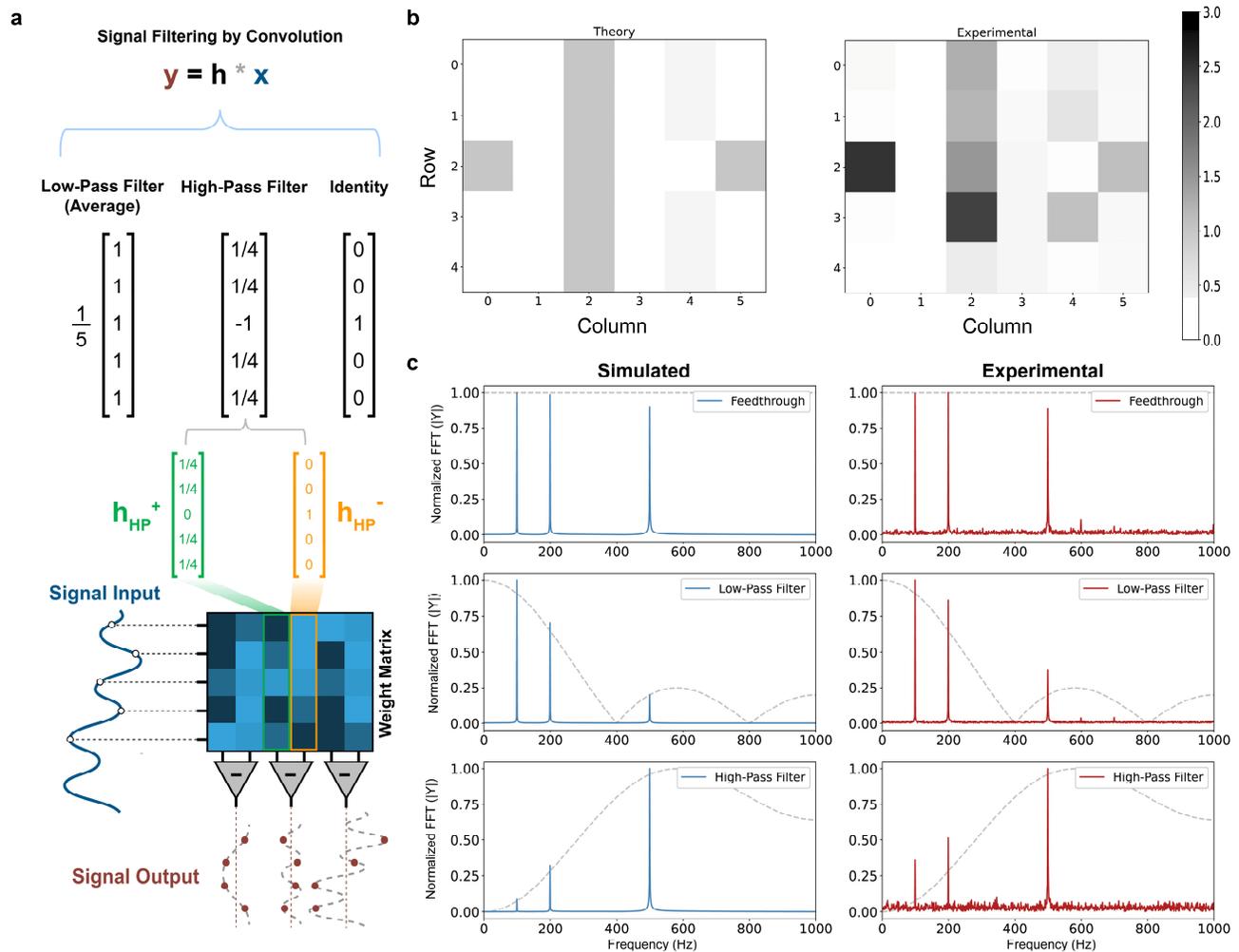

**Figure 4. Signal processing based on in-memory processing. a,** Description of convolution-based signal processing for different filters (low/high-pass filters and identity). y - the processed signal; x – the input signal; h – the filter kernel. The kernel is split between its positive and negative components; these values are proportionally transferred to the memory weights. The input signal is applied simultaneously to all memories and the difference between the output of two columns is the result of the processed signal for a given kernel. **b,** Comparison of the theoretical kernel weights mapping and the experimental weight transfer into the conductance of the memories. **c,** Comparison of the fast Fourier transform (FFT) of the simulated and experimental output signals after each kernel.